\documentclass[twocolumn,showpacs,aps,prl,letterpaper]{revtex4}
\usepackage{graphicx}
\usepackage{dcolumn}
\usepackage{amsmath}
\usepackage{epsfig}
\RequirePackage{xspace}

\usepackage{relsize}
\begin{document}

\title{
\large \bfseries \boldmath The continuum and interference effect
in $e^+e^-\to D^0\bar{D}^0,\; D^+D^-$ processes}

\author{Mao-Zhi Yang}\email{yangmz@mail.ihep.ac.cn}
\affiliation{ CCAST (World Laboratory), P.O.Box 8730, Beijing 100080, China  \\
 Institute of High Energy Physics, P.O.Box 918(4), Beijing  100049, China }
\date{\today}
\begin{abstract}
In the $e^+e^-$ annihilation processes $e^+e^-\to D^0\bar{D}^0,\;
D^+D^-$ near or above the threshold of $D\bar{D}$, there are not
only the resonance contribution $e^+e^-\to \psi (3770)\to
D^0\bar{D}^0,\; D^+D^-$, but also the continuum contribution
through virtual photon directly $e^+e^-\to \gamma^*\to
D^0\bar{D}^0,\; D^+D^-$. The amplitudes through virtual photon
directly and through resonance can interfere seriously. We
consider the continuum and interference effect in the $D\bar{D}$
production process in $e^+e^-$ annihilation. We find that the
effect is significant near and above the threshold of the
$D\bar{D}$ mesons.
\end{abstract}
\pacs{13.25.Gv, 13.66.Bc, 14.40.Gx}
% PACS, the Physics and Astronomy Classification Scheme.

\maketitle

The $e^+e^-$ annihilation processes $e^+e^-\to D^0\bar{D}^0$ and
$D^+D^-$ just above the $D\bar{D}$ threshold are important, they
occur significantly through the charmoniuum resonance
$\psi(3770)$. They can be used to determine the resonance
parameters of $\psi(3770)$, the mass $M$ and decay width $\Gamma$.
The resonance $\psi(3770)$ decays dominantly to $D\bar{D}$ final
state. Recently CLEO collaboration measured the difference between
the cross sections of $e^+e^-\to \psi (3770)\to \mbox{hadrons}$
and $e^+e^-\to \psi (3770)\to D\bar{D}$ at the center of mass
energy $E_{\mbox{cm}}=3.773\; \mbox{GeV}$. Their result is
$(-0.01\pm 0.08 ^{+0.41}_{-0.30})$ nb, which is consistent with
zero \cite{CLEO1}. This result indicates that the branching
fraction of $\psi(3770)\to$non-$D\bar{D}$ decay is small. In
addition, except for $J/\psi\pi^+\pi^-$ final state and some
radiative transition processes, no statistically significant
signals of non-$D\bar{D}$ decays of $\psi(3770)$ have been
detected \cite{PDG, CLEO2}. From the detected non-$D\bar{D}$ decay
modes of $\psi (3770)$ by experiment \cite{PDG}, one can estimate
that the summed branching fractions of $\psi (3770)$ to
non-$D\bar{D}$ final states is at most 2 or 3\%. Therefore about
98 or 97\% of $\psi (3770)$ resonances decay to $D\bar{D}$ final
states.

It is generally assumed that the process $e^+e^-\to D\bar{D}$ just
above the threshold of the $D\bar{D}$ is almost completely
contributed by $\psi(3770)$, while the continuum contribution of
virtual photon is not important.

BESII and CLEO-c have measured the cross sections of $e^+e^-\to
D^0\bar{D}^0$, and $D^+D^-$ at the center-of-mass energy
$\sqrt{s}=3.773\;\mbox{GeV}$ \cite{BESII,CLEO3}. Based on these
data, we show in this paper, if one accepts that the branching
ratio of $\psi(3770)\to D\bar{D}$ is as large as 97 or 98\%, to
understand the data of the cross section of $e^+e^-\to D\bar{D}$,
one has to consider both the continuum contribution of $e^+e^-\to
\gamma^* \to D\bar{D}$ and the interference effect between the
amplitudes of $e^+e^-\to \psi (3770)\to D\bar{D}$ and $e^+e^-\to
\gamma^* \to D\bar{D}$.

%\vspace{1cm}
\begin{figure}[h]
\epsfig{file=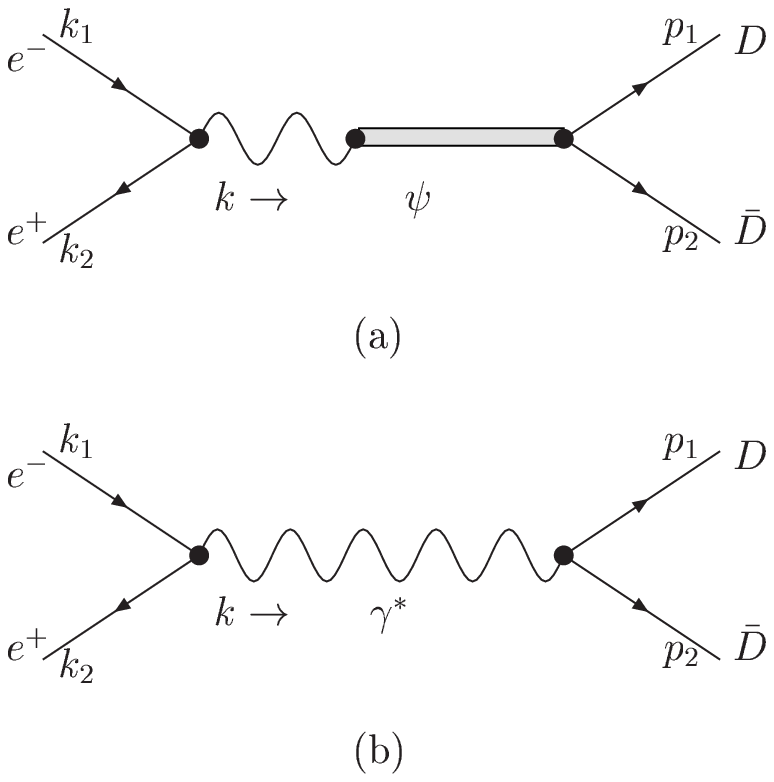,width=6cm,height=6cm}
\caption{Feynman diagrams for $e^+e^-\to D\bar{D}$ near the
resonance $\psi(3770)$.} \label{fig1}
\end{figure}

As shown in Fig.\ref{fig1}, the process of $e^+e^-$ annihilating
into $D\bar{D}$ can occur either through the intermediate
resonance $\psi(3770)$ or directly through the virtual photon. To
calculate these diagrams, one needs to know the coupling of
$\psi(3770)$ with $D\bar{D}$, the coupling of $\psi(3770)$ with
the virtual photon, and the form factor $F_{D\bar{D}}$ describing
the direct interaction of $D\bar{D}$ with the electromagnetic
current. The coupling $g_{\psi D\bar{D}}$ can be defined by
\begin{eqnarray}
&&\langle D(p_1)\bar{D}(p_2)|\psi(p)\rangle \nonumber\\
&=&-ig_{\psi D\bar{D}}\epsilon^{(\lambda)}\cdot
(p_1-p_2)(2\pi)^4\delta^4(p-p_1-p_2), \label{eq1}
\end{eqnarray}
where $\psi$ denotes $\psi(3770)$, $\epsilon^{(\lambda)}$ is the
polarization vector of $\psi(3770)$, and $\lambda$ stands for the
polarization state. The coupling of $\psi$ with the
electromagnetic current can be described by the following equation
\begin{equation}
\langle 0|\bar{c}\gamma_\mu c|\psi\rangle
=f_{\psi}m_{\psi}\epsilon^{(\lambda)}_\mu, \label{eq2}
\end{equation}
where $m_{\psi}$ is the mass of $\psi(3770)$, and $f_{\psi}$ its
decay constant, which can be determined by the measured decay
width of $\Gamma (\psi(3770)\to e^+e^-)$. In the following
analysis, we use $f_{\psi}=95.5 \mbox{MeV}$, which is extracted
from the measured decay width of $\psi(3770)\to e^+e^-$
\cite{PDG}. From eq.(\ref{eq2}), one can get the effective vertex
of the resonance $\psi$ and the electromagnetic current: $iQ_c e
f_{\psi}m_{\psi}\epsilon^{(\lambda)}_\mu$. The propagator of
$\psi$ resonance is taken as the Breit-Wigner form
\begin{equation}
\frac{i}{p^2-m_{\psi}^2+im_{\psi}\Gamma_T},
\end{equation}
here $\Gamma_T$ is the total decay width of $\psi(3770)$. Then one
can obtain the amplitude of Fig.{\ref{fig1}a
\begin{eqnarray}
T_a &=&ig_{\psi D\bar{D}}Q_ce^2f_{\psi}m_{\psi}(p_1-p_2)^\mu \nonumber\\
& \times & \frac{1}{s-m_{\psi}^2+im_{\psi}\Gamma_T}\frac{1}{s}
\bar{v}(k_2)\gamma_\mu u(k_1),\label{Ta}
\end{eqnarray}
where $s=(k_1+k_2)^2$, $u(k_1)$ and $v(k_2)$ are the Dirac spinors
for the electron and positron, respectively.

The coupling of the meson pair $D\bar{D}$ directly with the
electromagnetic current can be parameterized as the
electromagnetic form factor $F_{D\bar{D}}$
\begin{equation}
\langle
D(p_1)\bar{D}(p_2)|j^\mu_{\small\mbox{em}}|0\rangle_{|{\small\mbox{dir}}}
=F_{D\bar{D}}(q^2)(p_1-p_2)^\mu,
\end{equation}
where $j^\mu_{\mbox{em}}=Q_c \bar{c}\gamma^\mu c$, is the
electromagnetic current, $q=p_1+p_2$, and the subscript ``dir"
means the $D\bar{D}$ pair couples directly with the
electromagnetic current without the contribution of the resonance.
With this definition of the electromagnetic form factor, the
vertex of the virtual photon and $D\bar{D}$ is
$ieF_{D\bar{D}}(q^2)(p_1-p_2)^\mu$. Then the amplitude of
Fig.\ref{fig1}b can be obtained
\begin{equation}
T_b =-ie^2F_{D\bar{D}}(s)(p_1-p_2)^\mu \frac{1}{s}
\bar{v}(k_2)\gamma_\mu u(k_1).\label{Tb}
\end{equation}
Combining eqs.(\ref{Ta}) and (\ref{Tb}), one can obtain the total
amplitude of $e^+e^-\to D^0\bar{D}^0$ or $D^+D^-$
\begin{eqnarray}
T&=&ie^2\bar{v}(k_2)\gamma_\mu
u(k_1)(p_1-p_2)^\mu\frac{1}{s}\nonumber\\
&&\times [-F_{D\bar{D}}(s)+\frac{g_{\psi
D\bar{D}}Q_cf_{\psi}m_{\psi}}{s-m_{\psi}^2
+im_{\psi}\Gamma_T}e^{i\phi}],
\end{eqnarray}
where an arbitrary phase factor $e^{i\phi}$ is added, which should
be determined by experiment. With the above amplitude, one can
obtain the cross section of $e^+e^-\to D^0\bar{D}^0$ and/or
$D^+D^-$
\begin{eqnarray}
&&\sigma (e^+e^-\to D^0\bar{D}^0,\;D^+D^-)
=\frac{\pi}{3}\frac{(s-4m_D^2)^{3/2}}{s^{5/2}}\alpha^2\nonumber\\
&&\times |-F_{D\bar{D}}(s)+\frac{g_{\psi
D\bar{D}}Q_cf_{\psi}m_{\psi}}{s-m_{\psi}^2
+im_{\psi}\Gamma_T}e^{i\phi}|^2, \label{cross:section}
\end{eqnarray}
where $\alpha =1/137$, is the fine structure constant.

To proceed to the numerical analysis of the cross section of
$e^+e^-\to D\bar{D}$, one needs the knowledge of the coupling
$g_{\psi D\bar{D}}$ and the form factor $F_{D\bar{D}}(s)$. For the
coupling $g_{\psi D\bar{D}}$, we can estimate it from the
knowledge of the branching ratio of $\psi(3770)\to D^0\bar{D}^0$
and $\psi(3770)\to D^+D^-$. From eq.(\ref{eq1}) we can get the
branching ratio
\begin{equation}
Br(\psi(3770)\to D^0\bar{D}^0,D^+D^-)=\frac{g_{\psi
D\bar{D}}^2}{6\pi}\frac{|\vec{p}_D|^3}{m_{\psi}^2}\frac{1}{\Gamma_T},
\end{equation}
where $\vec{p}_D$ is the three-momentum of the $D$ meson in
$\psi(3770)$ decay. The isospin violation effect for $g_{\psi
D\bar{D}}$ is not considered here, since it is only at one or two
percentage level. We take the mass and the total width of
$\psi(3770)$ as $m_{\psi}=3771.1\mbox{MeV}$,
$\Gamma_T=23.0\mbox{MeV}$ \cite{PDG}, and the total branching
fraction of $\psi(3770)$ decays to $ D^0\bar{D}^0$ and $D^+D^-$ to
be $Br(\psi(3770)\to D\bar{D})=97\%$ in the numerical estimation,
which is relevant to take $g_{\psi D\bar{D}}=12.7$. The value of
$g_{\psi D\bar{D}}$ and the results for the branching ratios of
$\psi(3770)\to D\bar{D}$ are listed in Table \ref{table1}. The
ratio of $\frac{Br(D^0\bar{D}^0) }{Br( D^+D^-)}=1.47$, which is
within the error of the experimental value: $2.43\pm 1.50\pm 0.43$
\cite{Belle}.

\begin{table}\caption{Branching ratio of $Br(\psi(3770)\to
D\bar{D})$ and the coupling $g_{\psi D\bar{D}}$.}\label{table1}
\begin{tabular}{c|c|c|c|c}\hline
$g_{\psi D\bar{D}}$& $ Br( D\bar{D})$   & $Br( D^0\bar{D}^0) $
&$Br( D^+D^-)$ &
$\frac{Br(D^0\bar{D}^0) }{Br( D^+D^-)}$\\
\hline  12.7 &97.4\% &58.0\% & 39.4\% & 1.47 \\ \hline
\end{tabular}
\end{table}

Now we consider the cross section of $e^+e^-\to D^0\bar{D}^0$ and
$D^+D^-$ with $g_{\psi D\bar{D}}=12.7$. At first we calculate the
cross section without the continuum contribution, i.e., taking the
form factor $F_{D\bar{D}}(s)=0$ in eq. (\ref{cross:section}). The
results are listed in Table \ref{table2}. Both ClEO-c and BESII
have measured the cross sections of $e^+e^-\to D^0\bar{D}^0$ and
$D^+D^-$ at $\sqrt{s}=3.773\;\mbox{GeV}$. Their results are in
good agreement with each other. The observed cross sections are
\[
\begin{array}{ll}
\left\{\begin{array}{ll}
\sigma^{obs}_{D^0\bar{D}^0}=(3.58\pm 0.09\pm 0.31)\; \mbox{nb}&\\
\sigma^{obs}_{D^+D^-}=(2.56\pm 0.08\pm 0.26)\; \mbox{nb}&
\end{array}\right.   & \mbox{BESII \cite{BESII}} \\[4mm]
\left\{\begin{array}{ll}
\sigma^{obs}_{D^0\bar{D}^0}=(3.60\pm 0.07^{+0.07}_{-0.05})\; \mbox{nb}&\\
\sigma^{obs}_{D^+D^-}=(2.79\pm 0.07^{+0.10}_{-0.04})\; \mbox{nb}&
\end{array} \right.  &  \mbox{CLEO-c \cite{CLEO3}}
\end{array}
\]
BESII has also given the cross sections with the initial state
radiative corrections. This result can be compared directly with
our tree-level calculation. The comparison can be found in Table
\ref{table2}. It shows that, without the continuum contribution,
both of the results of the cross sections of $e^+e^-\to
D^0\bar{D}^0$ and $D^+D^-$ are seriously larger than the
experimental data at $\sqrt{s}=3.773\;\mbox{GeV}$. To improve the
theoretical prediction, one choice is to include the contribution
of the direct virtual photon in the $e^+e^-\to D\bar{D}$ process.

\begin{table}[h]
\caption{The cross sections of  $e^+e^-\to D^0\bar{D}^0$ and
$D^+D^-$ at $\sqrt{s}=3.773\;\mbox{GeV}$ without the direct
virtual photon contribution, which are compared with experimental
data. The experimental data are taken from
Ref.\cite{BESII}.}\label{table2}
\begin{tabular}{ccc}\hline
 & The calculation & Exp. data \\ \hline
 $\sigma_{D^0\bar{D}^0}$ & 6.5 nb & $(4.6\pm 0.12\pm
 0.45)~\mbox{nb}$ \\
 $\sigma_{D^+D^-}$ & 4.5 nb & $(3.29\pm 0.10\pm
 0.37)~\mbox{nb}$  \\ \hline
\end{tabular}
\end{table}

For considering the contribution of the intermediate virtual
photon, we assume the electromagnetic form factor of $D\bar{D}$ to
be the following form
\begin{equation}
F_{D\bar{D}}(s)=\frac{m_{\psi}^2F_0}{s}. \label{form}
\end{equation}
This behavior of $s$-dependence is similar to the form factors of
light mesons \cite{s:dependence, wmy}.

\begin{figure}[h]
\scalebox{1.1}{\epsfig{file=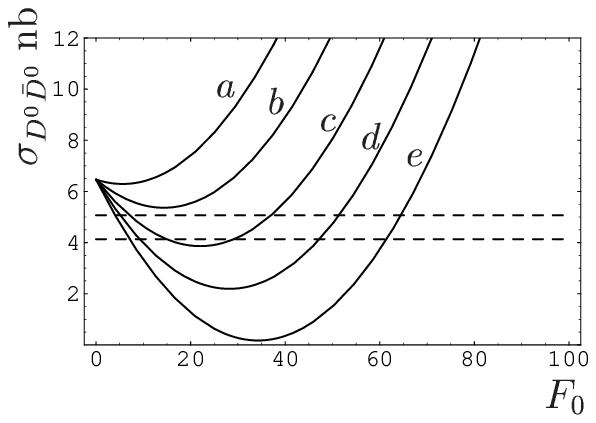}} \caption{\small The cross
section of $e^+e^-\to D^0\bar{D}^0$ at
$\sqrt{s}=3.773\;\mbox{GeV}$, varying with the parameter $F_0$ and
the relative phase $\phi$. The plots $a$, $b$, $c$, $d$ and $e$
are for $\phi=0$, $\pi/12$, $\pi/6$, $\pi/4$ and $\pi/2$,
respectively.}\label{fig2}
\end{figure}

Now substitute eq.(\ref{form}) into eq.(\ref{cross:section}), one
can calculate the cross section depending on the center-of-mass
energy $\sqrt{s}$, provided one knows the values of the parameter
$F_0$ in eq.(\ref{form}) and the relative phase $\phi$ between the
resonance and the direct virtual photon contribution in eq.
(\ref{cross:section}). To estimate the allowed values for $F_0$
and $\phi$, we plot the cross section of $e^+e^-\to D^0\bar{D}^0$
in Fig.\ref{fig2} by varying the values of $F_0$ and $\phi$. The
two dashed horizontal lines stand for the experimental result with
the upper and lower errors \cite{BESII}. The curves between this
two dashed lines are allowed by the experimental data. Similar
plots can be obtained for the cross section of $e^+e^-\to D^+D^-$.
We find the common parameter space of $(F_0,\phi )$ exist, which
can accommodate the data for both of the two annihilation
processes $e^+e^-\to D^0\bar{D}^0$ and $D^+D^-$. Some illustrative
results are given in Table \ref{table3}.

\begin{table}[h]
\caption{The numerical results for the cross sections of
$e^+e^-\to D^0\bar{D}^0$ and $D^+D^-$ at
$\sqrt{s}=3.773\;\mbox{GeV}$ with some values of $(F_0,\phi
)$.}\label{table3}
\begin{tabular}{ccccc}\hline
 $(F_0,\phi )$ & $(8.0,\pi/6)$& $(6.0,\pi/4)$ & $(5.0,\pi/2)$& Exp. data \cite{BESII} \\ \hline
 $\sigma_{D^0\bar{D}^0}$ & 4.9 nb & 4.8 nb& 4.8 nb &$(4.6\pm 0.12\pm
 0.45)\mbox{nb}$ \\
 $\sigma_{D^+D^-}$ & 3.4 nb & 3.4 nb & 3.3 nb &$(3.29\pm 0.10\pm
 0.37)\mbox{nb}$  \\ \hline
\end{tabular}
\end{table}

The $s$-dependence of the cross section of $e^+e^-\to
D^0\bar{D}^0$ are shown in Fig. \ref{fig3}, by taking
$(F_0,\phi)=(8.0,\pi/6)$, $(6.0,\pi/4)$ and $(5.0,\pi/2)$. Both of
these sets of parameters can give the correct prediction at
$\sqrt{s}=3.773\;\mbox{GeV}$. The two dashed horizontal lines are
the experimental bound at $\sqrt{s}=3.773\;\mbox{GeV}$. The solid
curve is for $(F_0,\phi)=(5.0,\pi/2)$, the dotted one for
$(6.0,\pi/4)$, and the dashed one for $(8.0,\pi/6)$. The curves
for the cross section changing with the center-of-mass energy
$\sqrt{s}$ are very sensitive to the form-factor parameter $F_0$
and the relative phase $\phi$. With precisely measured cross
section of $\sigma_{D\bar{D}}(s)$ at enough different points of
$\sqrt{s}$, these parameters can be possibly fitted well. This
measurement can be performed in BESIII and CLEO-c.

\begin{figure}[h]
\scalebox{1.1}{\epsfig{file=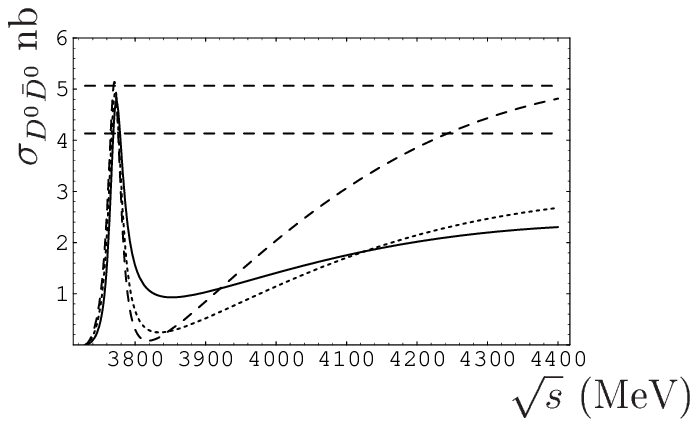}} \caption{\small The cross
section of $e^+e^-\to D^0\bar{D}^0$ changing with the
center-of-mass energy $\sqrt{s}$. The solid curve is for
$(F_0,\phi)=(5.0,\pi/2)$, the dotted one for $(6.0,\pi/4)$, and
the dashed for $(8.0,\pi/6)$.}\label{fig3}
\end{figure}

How important is the continuum contribution in the resonance
region in the process of $e^+e^-\to D\bar{D}$? The plot for each
contribution by taking $(F_0,\phi)=(5.0,\pi/2)$ is shown in
Fig.\ref{fig4}, where the dotted curve is the single contribution
of the resonance $\psi(3770)$, the dashed curve is the single
contribution of the virtual photon, and the solid one is the total
contribution. At $\sqrt{s}=3.773\;\mbox{GeV}$, the single
contribution of the resonance $\psi(3770)$ is 6.5 nb, the
continuum contribution is 0.13 nb, while the combined contribution
is 4.8 nb. This shows that the continuum contribution can change
the final cross section as large as 30\%. The interference effect
is destructive. In the resonance region, the peak for the curve of
the cross section is slightly shifted, and seriously lowered by
the continuum virtual photon contribution.

\begin{figure}[h]
\scalebox{1.1}{\epsfig{file=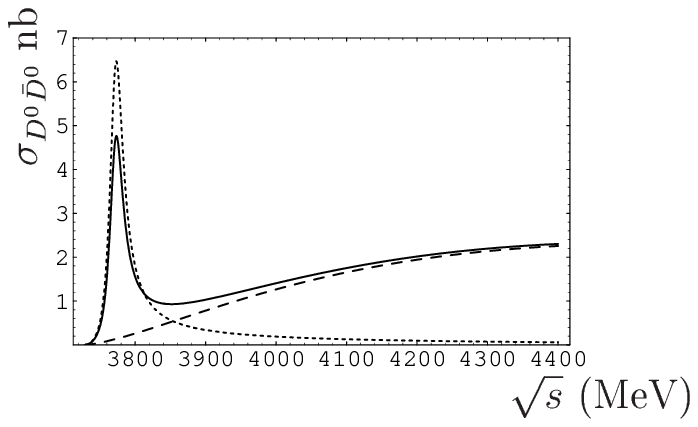}} \caption{\small The cross
section of $e^+e^-\to D^0\bar{D}^0$ changing with the
center-of-mass energy $\sqrt{s}$. The solid curve is  the total
contribution of the virtual photon and the resonance $\psi(3770)$,
where $(F_0,\phi)=(5.0,\pi/2)$ is taken, while the dotted one is
for the case without the continuum virtual photon contribution,
and the dashed one is only the virtual photon
contribution.}\label{fig4}
\end{figure}

In the numerical analysis, we did not include the possible effect
of the lower resonance $\psi (3686)$, which is just under the
threshold of $D\bar{D}$, because we do not know any information
about its coupling with $D\bar{D}$. When performing the
experimental analysis, the contribution of this resonance can be
added into the formula of eq.(\ref{cross:section}). With large
amount of experimental data about the cross section of $e^+e^-\to
D\bar{D}$ at different point of $\sqrt{s}$, it is possible to
detect whether the effect of the resonance $\psi(3686)$ is
important. In general eq. (\ref{cross:section}) can be generalized
as
\begin{eqnarray}
&&\sigma_{D\bar{D}}
=\frac{\pi}{3}\frac{(s-4m_D^2)^{3/2}}{s^{5/2}}\alpha^2\nonumber\\
&&\times |-F_{D\bar{D}}(s)+\sum_{i}\frac{g_{\psi_i
D\bar{D}}Q_cf_{\psi_i}m_{\psi_i}}{s-m_{\psi_i}^2
+im_{\psi_i}\Gamma_{i}}e^{i\phi_i}|^2, \label{cross:section2}
\end{eqnarray}
where the summation is over all the possible resonances which can
couple to $D\bar{D}$. However, in practice, one need only consider
the resonances which give large contribution.

It is still possible to understand the experimental data of the
cross section of $e^+e^-\to D\bar{D}$ theoretically by decreasing
the coupling of $\psi(3770)$ with $D\bar{D}$, $g_{\psi D\bar{D}}$.
However, this will predict a  drastically smaller branching ratio
of $\psi(3770)\to D\bar{D}$, then give a larger branching ratio
for the non-$D\bar{D}$ decays of $\psi(3770)$. From the analysis
of Ref \cite{wang} it is concluded that the branching ratio of the
charmless non-$D\bar{D}$ decays of $\psi(3770)$ may be as large as
13\% or 18\%. Such a large branching fraction for non-$D\bar{D}$
decays of $\psi(3770)$ is not favored by the recent measurement of
CLEO collaboration \cite{CLEO1}. Without considering the direct
virtual photon contribution, it would be a puzzle for
understanding consistently the measured cross section of
$e^+e^-\to D\bar{D}$ at the resonance region and the experimental
data about the non-$D\bar{D}$ decays of $\psi(3770)$ \cite{PDG,
CLEO2}. Now the data can be understood consistently by considering
the continuum and interference effect caused by the direct virtual
photon in the annihilation process of $e^+e^-\to D\bar{D}$.

\vspace{1cm}

\begin{figure}[ht]
\scalebox{0.65}{\epsfig{file=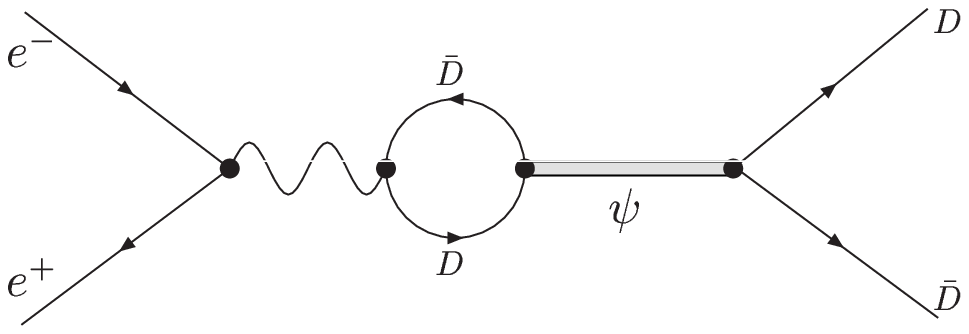}} \caption{\small The
$D\bar{D}$ loop diagram for $e^+e^-\to D\bar{D}$ scattering.
}\label{fig5}
\end{figure}

One more remark is in the following. The re-scattering
contribution of $e^+e^-\to \gamma^*\to D\bar{D}\to \psi\to
D\bar{D}$ shown in Fig.\ref{fig5} is not considered separately in
this work, because this diagram has been absorbed into
Fig.\ref{fig1} (a) by using the effective coupling of $\psi(3770)$
with the electromagnetic current extracted from the measured decay
width of $\psi(3770)\to e^+e^-$. The $D\bar{D}$ loop gives a kind
of long-distance contribution to the effective vertex of the
electromagnetic field and the resonance $\psi(3770)$. When taking
the value for this effective vertex extracted from experimental
data, all short and long-distance contributions are included.
Therefore, it is not necessary to consider Fig.\ref{fig5} again.

In summary, we have considered the continuum and interference
effect in the process of $e^+e^-\to D\bar{D}$ near and above the
threshold of $D\bar{D}$. We find the effect of the direct virtual
photon is crucial. By including the contribution of the direct
virtual photon in $e^+e^-\to D\bar{D}$, the recent experimental
data about the non-$D\bar{D}$ decays of $\psi(3770)$ and the
measured cross section of $e^+e^-\to D\bar{D}$ can be understood
consistently.

%%%%%%%%%%%%%%%%%%%%%%%%%%%%%%%%%%%%%%%%%%%%%%%%%%%%%%%%%%%%%%%%%%%%%%%%%
% ACKNOWLEDGMENTS
%%%%%%%%%%%%%%%%%%%%%%%%%%%%%%%%%%%%%%%%%%%%%%%%%%%%%%%%%%%%%%%%%%%%%%%%%
The author would like to thank Hai-Bo Li for stimulating
discussions. This work is supported in part by the National
Natural Science Foundation of China under contract Nos. 10205017,
10575108, and the Knowledge Innovation Project of CAS under
contract U-530 (IHEP).

%%%%%%%%%%%%%%%%%%%%%%%%%%%%%%%%%%%%%%%%%%%%%%%%%%%%%%%%%%%%%%%%%%%%%%%%%

\end{document}